\documentclass[aps,pra,preprint,superscriptaddress,showpacs]{revtex4-2}

\usepackage{amsmath}
\usepackage{amssymb}
\usepackage{graphicx}
\usepackage{bm}

\begin{document}

\title{Eigenvalue-by-Eigenvalue Comparison of a Sierra--Rodr\'iguez-Laguna-Type
Spectrum with the Riemann Zeros}

\author{Mi-Ra Hwang}
\affiliation{Department of Electronic Engineering, Kyungnam University, Changwon, 631-701, Korea}
\author{Eylee Jung}
\affiliation{Department of Electronic Engineering, Kyungnam University, Changwon, 631-701, Korea}
\author{MuSeong Kim}
\affiliation{Pharos iBio Co., Ltd. Head Office: \#1408, 38, Heungan-daero 427beon-gil, Dongan-gu, Anyang, 14059, Korea}
\author{DaeKil Park}
\email{dkpark@kyungnam.ac.kr}
\affiliation{Department of Electronic Engineering, Kyungnam University, Changwon, 631-701, Korea}
\affiliation{Department of Physics, Kyungnam University, Changwon, 631-701, Korea}

\begin{abstract}
We study the self-adjoint extension of the Sierra--Rodr\'iguez-Laguna (SR) $H=xp$-type
Hamiltonian $\widehat H_{SR}$. Its discrete spectrum is fixed by the equation
$\mathrm{Re}\!\left[e^{-i\theta/2}K_{1/2+iE/2}(2\pi)\right]=0$. We solve this equation
numerically for $\theta=1.417\pi$ and obtain the first 606 eigenvalues $E_n$. We compare
them, one by one, with the ordinates $\gamma_n$ of the first 606 nontrivial zeros of the
Riemann zeta function. Using the steepest-descent (saddle-point) method for the modified
Bessel function, together with the Riemann--von Mangoldt formula, we derive a closed-form
prediction for $E_n-\gamma_n$ in terms of the Lambert-$W$ function:
$E_n-\gamma_n\simeq(\theta-\pi/4)/[1+W(n/e)]$, which decays only like $1/\ln n$. Here
the $n$-th numerical root corresponds to the $(n-2)$-th root of the asymptotic
quantization condition. The leading-order phase of $K_{1/2+iE/2}(2\pi)$ agrees with a
known rigorous asymptotic formula for modified Bessel functions of large imaginary order.
The prediction is checked against the first $10^5$ tabulated zeta zeros (A.~Odlyzko): the
measured coefficient of $1/[1+W(n/e)]$ is $3.666$ (mean) and $3.667$ (median) for
$n\ge10^3$, in agreement with $\theta-\pi/4=3.666$. The scatter of $E_n-\gamma_n$ around
this smooth curve comes from the fluctuating term $S(T)$ of the zero-counting function,
which the spectrum $E_n$ does not contain.
\end{abstract}

\maketitle

\section{Introduction}

The Hilbert--P\'olya conjecture says that the nontrivial zeros $\tfrac12+i\gamma_n$ of
the Riemann zeta function might be the eigenvalues $\gamma_n$ of some self-adjoint
operator. If true, this would give a simple reason why the Riemann Hypothesis is true.
Many people have looked for such an operator.

Berry and Keating (BK)~\cite{BK} proposed the simplest possible candidate: the classical
Hamiltonian $H_{BK}=xp$. Its periodic-orbit sum looks similar to the explicit formula
for the prime-counting function, and a rough quantization of $H_{BK}$ reproduces the
mean density of the zeros, $\overline{N}(T)\sim\tfrac{T}{2\pi}\ln\tfrac{T}{2\pi e}$.
However, $H_{BK}$ has a continuous spectrum, not a discrete one, so it cannot give the
individual zeros. A later, more careful study treated $H=(x+\ell_x^2/x)(p+\ell_p^2/p)$ on the half line $x>0$ and quantized it~\cite{BK2011}; the counting function of related models was also discussed in Ref.~\cite{Sierra12}.

Sierra and Rodr\'iguez-Laguna (SR)~\cite{SR} later modified $H_{BK}$ to
\begin{equation}
H_{SR}=x\left(p+\frac{\ell_p^2}{p}\right),\qquad x\ge \ell_x ,
\end{equation}
by adding a term $\ell_p^2/p$ and placing a hard wall at $x=\ell_x$. Unlike $H_{BK}$,
this Hamiltonian has a classically periodic orbit. Once we choose a boundary
condition at $x=\ell_x$, the quantum problem gets a genuinely discrete spectrum. SR
showed numerically that this spectrum reproduces the known statistics of the Riemann
zeros (spacing distribution, counting function). In this paper we ask a more direct
question: can the individual eigenvalues $E_n$ be matched, one by one, to the individual
zeros $\gamma_n$? And if so, how fast does $E_n-\gamma_n\to0$ as $n\to\infty$?

The Schr\"odinger equation of the model is
\begin{equation}
\widehat H_{SR}\psi(x)=E\psi(x),
\end{equation}
with $\widehat H_{SR}=x^{1/2}\!\left(\hat p+\ell_p^2/\hat p\right)x^{1/2}$. This operator
becomes self-adjoint, and its spectrum becomes discrete, once we impose
$\langle\psi_1|\widehat H_{SR}\psi_2\rangle=\langle\widehat H_{SR}\psi_1|\psi_2\rangle$.
This condition reduces to a single transcendental equation,
\begin{equation}
\Xi_{\widehat H_{SR}}\equiv \mathrm{Re}\!\left[e^{-i\theta/2}K_{1/2+iE/2}(2\pi)\right]=0,
\label{eq:xi}
\end{equation}
where $K_\nu(z)$ is the modified Bessel function, $\theta$ is a free (self-adjoint
extension) parameter, and $2\pi\hbar=\ell_x\ell_p$. We set $\hbar=1$ from here on.

\section{Numerical spectrum versus the zeta zeros}

We solved Eq.~\eqref{eq:xi} numerically with $\theta=1.417\pi$ and obtained
$E_1,\dots,E_{606}$. (One more apparent root, near $E\approx947.5$, turned out to be a
numerical artifact --- most likely two very close roots merged into one by the
root-finder --- and was removed; see the check described near the end of
Sec.~\ref{sec:num}.) Figure 1(a) shows $E_n$ together with $\gamma_n$, the
$n$-th zero ordinate on the critical line $1/2+i\gamma_n$ as a red and blue dots; the two sequences look almost
identical. Figure 1(b) shows $E_n-\gamma_n$. Three values are negative:
$E_{289}-\gamma_{289}=-0.102333$, $E_{519}-\gamma_{519}=-0.0197868$, and
$E_{568}-\gamma_{568}=-0.0727416$. For all other $n$, $E_n>\gamma_n$. The largest
difference is $|E_n-\gamma_n|\approx2.0$, and the difference seems to shrink slowly as
$n$ grows. Below we show that $E_n-\gamma_n\propto 1/\ln n\to0$ as $n\to\infty$. The labeling of the roots matters here: $E_n$ is the $n$-th root found numerically, counted from the smallest one.


\begin{figure}[ht!]
\begin{center}
\includegraphics[height=5.2cm]{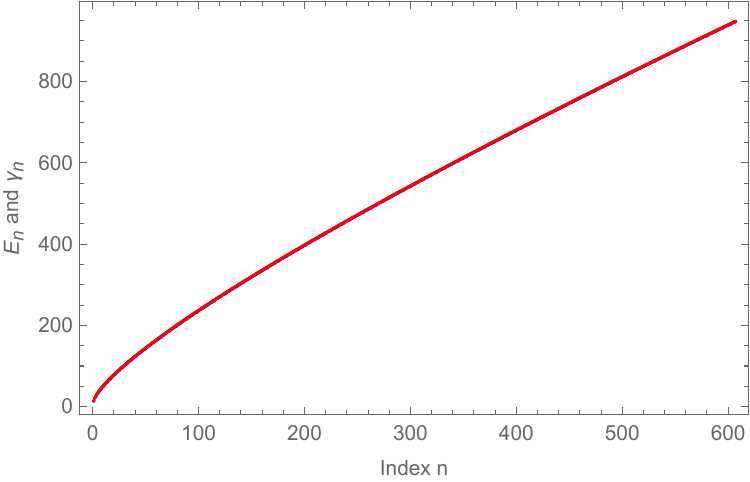} 
\includegraphics[height=5.2cm]{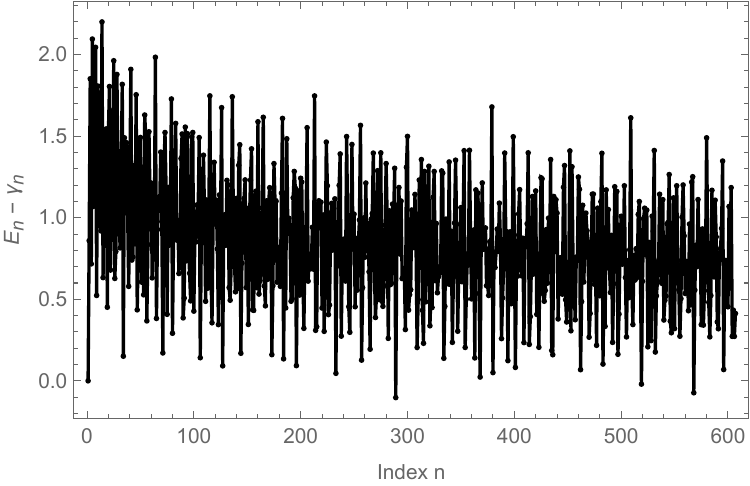}

\caption[fig1]{(Color online) (a) $E_n$ (red) and $\gamma_n$ (blue) for $n=1,\dots,606$.
(b) $E_n-\gamma_n$. }
\label{fig:fig1}
\end{center}
\end{figure}

\section{Large-$E$ asymptotics of $K_{1/2+iE/2}(2\pi)$}
\label{sec:saddle}

To study Eq.~\eqref{eq:xi} for large $E$, we use the integral form
\begin{equation}
K_\mu(x)=\frac12\int_{-\infty}^{\infty} e^{\phi(t)}\,dt,\qquad
\phi(t)=-x\cosh t+\mu t,
\label{eq:intrep}
\end{equation}
with $x=2\pi$ and $\mu=\tfrac12+\tfrac{iE}{2}$, and apply the steepest-descent method.
The saddle-point condition $\phi'(t_0)=0$ gives
$e^{t_0}-e^{-t_0}=\tfrac1{2\pi}(1+iE)\approx \tfrac{iE}{2\pi}$. Because $\sinh$ has
period $2\pi i$, this equation formally has infinitely many solutions,
$t_0+2\pi ik$ and $t_0'+2\pi ik$ ($k\in\mathbb Z$), where
\begin{equation}
t_0\approx a_0+i\frac\pi2,\qquad t_0'\approx -a_0+i\frac\pi2,\qquad
a_0=\ln\frac{E}{2\pi}.
\end{equation}
Only $t_0$ and $t_0'$ --- the two saddles closest to the original path
$\mathrm{Im}\,t=0$ --- can actually be reached by smoothly deforming that path. The
other saddles $t_0+2\pi ik$ ($k\ne0$) satisfy $\phi'=0$ too, and formally have
$\mathrm{Re}[\phi(t_0+2\pi ik)]=\mathrm{Re}[\phi(t_0)]-\pi kE$, which can be much larger
than $\mathrm{Re}[\phi(t_0)]$. But they are not connected to the original path, so they
do not contribute. We can check this: if one of them did contribute, then
$K_{1/2+iE/2}(2\pi)$ would blow up as $E\to\infty$ instead of going to zero, which
contradicts the known behavior of $K_\mu(x)$.

Expanding $\phi(t)=\phi(t_0)+\tfrac12\phi''(t_0)(t-t_0)^2+\cdots$ gives
\begin{align}
\phi(t_0)&\approx -\left(\frac{\pi E}{4}-\frac12\ln\frac{E}{2\pi}\right)
+i\left(\frac{E}{2}\ln\frac{E}{2\pi e}+\frac\pi4\right),\notag\\
\phi''(t_0)&\approx-\frac{iE}{2}.
\label{eq:phit0}
\end{align}
Let $a\equiv-\tfrac12\phi''(t_0)=iE/4$. The Gaussian integral then gives
\begin{equation}
\int_{-\infty}^\infty e^{\phi(t)}dt = e^{\phi(t_0)}\sqrt{\frac{\pi}{a}}
\left[1+\sum_{m\ge2}C_m\frac{(2m-1)!!}{(2a)^m}\right],
\label{eq:gauss}
\end{equation}
where each $C_m$ is built from $\phi^{(n)}(t_0)=\phi''(t_0)=-iE/2$ for $n\ge3$. A direct
check shows $C_m/a^m=O(E^{-1})$ for every $m\ge2$. For example, $C_2\propto
\phi^{(4)}(t_0)\propto E$, so $C_2a^{-2}\propto E^{-1}$; and $C_3$, built from
$\phi^{(6)}(t_0)$ and $[\phi^{(3)}(t_0)]^2$ (both $\propto E^2$), gives
$C_3a^{-3}\propto E^{-1}$ as well. So the Gaussian approximation is well controlled for
large $E$. Keeping only the dominant saddle $t_0$,
\begin{equation}
\int_{-\infty}^\infty e^{\phi(t)}dt=\sqrt2\,e^{-\pi E/4}
e^{i\frac{E}{2}\ln\frac{E}{2\pi e}}\left[1+O(E^{-1})\right].
\label{eq:leading}
\end{equation}
The other saddle, $t_0'$, contributes
$e^{-\pi E/4}\,2\sqrt{2\pi/E}\,e^{i(\pi/2-\Phi)}[1+O(E^{-1})]$, with
$\Phi\equiv\tfrac E2\ln\tfrac E{2\pi e}$. This term has the {\it same} exponential
factor $e^{-\pi E/4}$ as the main term; it is smaller only by the power-law factor
$\sqrt{8\pi/E}=O(E^{-1/2})$, not exponentially smaller. Its phase $\pi/2-2\Phi$ is fixed once $E$ is given, so it gives a smooth shift of each
root of order $E^{-1/2}$ (about $0.01$ near $E\simeq10^3$). Because it is $O(E^{-1/2})$, it
cannot explain a steady $O(1)$ phase shift, and it is small compared with $d_n = E_n - \gamma_n$ itself
for the range studied here. So, to the leading order we need,
\begin{equation}
K_{1/2+iE/2}(2\pi)\ \propto\ \exp\!\left[i\frac{E}{2}\ln\frac{E}{2\pi e}\right].
\label{eq:Kasym}
\end{equation}

\subsection{Check against the exact asymptotics for large imaginary order}

The case studied here --- $K_\nu(x)$ with $x$ fixed and order $\nu=r+it$, $t\to\infty$
--- is already known in the literature. It was solved rigorously by
Temme~\cite{Temme94}, building on earlier work by Balogh~\cite{Balogh67}; see also
Booker, Str\"ombergsson, and Then~\cite{BST13}, and a recent simpler derivation by
Tseng~\cite{Tseng23}. Writing $t=x\cosh\mu$ (here $x=2\pi$, $t=E/2$, so
$\mu\to a_0=\ln(E/2\pi)$ for large $E$), the exact result is
\begin{align}
K_{r+it}(x)={}&\sqrt{\frac{2\pi}{x\sinh\mu}}\,e^{-\frac{\pi x}2\cosh\mu+i\frac{r\pi}2}
\Big[\cosh(r\mu)\sin\!\big(\tfrac\pi4-\chi\big)\notag\\
&-i\sinh(r\mu)\cos\!\big(\tfrac\pi4-\chi\big)\Big]\big(1+o(1)\big),
\label{eq:temme}
\end{align}
with $\chi=x(\sinh\mu-\mu\cosh\mu)$. Set $r=1/2$ and let $\mu\to\infty$, so
$\cosh(\mu/2)\sim\sinh(\mu/2)\sim\tfrac12 e^{\mu/2}$. The bracket in Eq.~\eqref{eq:temme}
then becomes $\tfrac12 e^{\mu/2}(-i)e^{i(\pi/4-\chi)}$. Multiplying this by the prefactor
$e^{ir\pi/2}=e^{i\pi/4}$ gives a total phase
$e^{i\pi/4}\cdot e^{-i\pi/2}\cdot e^{i\pi/4}\cdot e^{-i\chi}=e^{-i\chi}$: the two
$\pi/4$ terms cancel exactly, and only $e^{-i\chi}$ is left. Using
$\chi\approx-\tfrac E2\ln\tfrac E{2\pi e}+O(E^{-1})$ (from expanding
$\mu=\mathrm{arccosh}(E/4\pi)$ for large $E$), this gives back Eq.~\eqref{eq:Kasym}
exactly, with no extra $O(1)$ phase. This match between our simple saddle-point
calculation and the rigorous result of Refs.~\cite{Temme94,Balogh67,BST13,Tseng23} is
strong evidence that no hidden phase is left on the Bessel-function side. Any leftover
mismatch between Eq.~\eqref{eq:quant} below and the true spectrum must come from
the labeling of the roots or from the zeta side.

\section{Quantization condition and the Lambert-$W$ solution}

Putting Eq.~\eqref{eq:Kasym} into Eq.~\eqref{eq:xi}, the quantization condition becomes
\begin{equation}
\cos\!\left(\frac{E}{2}\ln\frac{E}{2\pi e}-\frac\theta2\right)=0,
\label{eq:cos}
\end{equation}
with solutions labeled by an integer $m$,
\begin{equation}
\frac E2\ln\frac{E}{2\pi e}-\frac\theta2=\left(m+\frac12\right)\pi.
\label{eq:quant}
\end{equation}
This can be solved in closed form with the Lambert-$W$ function:
\begin{equation}
E^{\rm asym}_m = g\!\left(m+\frac12+\frac{\theta}{2\pi}\right),\qquad
g(x)\equiv\frac{2\pi x}{W(x/e)} .
\label{eq:Em}
\end{equation}
Equation~\eqref{eq:quant} is valid only for large $E$, so the integer $m$ is not
automatically the same as the number $n$ of the root counted from the bottom of the
spectrum. We fixed the relation by comparing with the exact roots of Eq.~\eqref{eq:xi}.
The $n$-th numerical root corresponds to $m=n-2$, that is,
\begin{equation}
E_n = g\!\left(n-\frac32+\frac{\theta}{2\pi}\right).
\label{eq:En}
\end{equation}
For example, for $n=600$ the direct numerical solution of Eq.~\eqref{eq:xi} gives
$E_{600}=939.475$, while Eq.~\eqref{eq:En} with $\theta=1.417\pi$ gives $939.484$. The
difference is only $0.009$. If we used $m=n$ instead, we would get $941.993$, which is
off by $2.5$, more than one level spacing ($\Delta E\simeq 2\pi/\ln(E/2\pi)\simeq1.3$). The
relation $m=n-2$ was checked in the same way for many other $n$ in the range
$200\le n\le600$.

\section{Comparison with the Riemann--von Mangoldt formula}
\label{sec:RvM}

For large $T$, the Riemann--von Mangoldt formula~\cite{Edwards,Titchmarsh} gives the
number of zeta zeros below $T$ as
\begin{equation}
N(T)=\frac{\theta(T)}{\pi}+1+S(T),\qquad
\theta(T)=\frac T2\ln\frac{T}{2\pi e}-\frac\pi8+O(T^{-1}),
\label{eq:NT}
\end{equation}
where $S(T)=\tfrac1\pi\arg\zeta(\tfrac12+iT)$, with the argument tracked continuously
along the path $2\to2+iT\to\tfrac12+iT$. Equation~\eqref{eq:NT} holds when $T$ is not
the ordinate of a zero. Exactly at such a point, $N(T)$ jumps:
$N(\gamma_n^-)=n-1\to N(\gamma_n^+)=n$ (we use the usual right-continuous definition,
$N(T)=\#\{0<\gamma\le T\}$), while $\theta(T)$ stays smooth. By the usual convention,
$S(T)$ at such a jump is taken to be the average of its two one-sided limits. If we use
this averaged $S(\gamma_n)$ in Eq.~\eqref{eq:NT}, we do {\it not} get the
right-continuous value $N(\gamma_n)=n$; we get the midpoint of the jump instead:
\begin{equation}
\frac{\theta(\gamma_n)}{\pi}+1+S(\gamma_n) = \frac{N(\gamma_n^-)+N(\gamma_n^+)}{2}
= n-\frac12 .
\label{eq:midpoint}
\end{equation}
Dropping $S(\gamma_n)$, since its long-run average is zero~\cite{Littlewood}, and
solving Eq.~\eqref{eq:midpoint} for $\gamma_n$ gives $\theta(\gamma_n)\approx
(n-3/2)\pi$, or
\begin{equation}
\gamma_n \sim g\!\left(n-\frac{11}{8}\right),
\label{eq:gamman}
\end{equation}
{\it not} $g(n-7/8)$, which is what a naive use of $N(\gamma_n)=n$ in
Eq.~\eqref{eq:NT} would give. As a rough check, even at $n=1$ --- far outside the range
where Eq.~\eqref{eq:gamman} should really apply --- the corrected formula $g(1-11/8)$
gives $14.52$, close to the true value $\gamma_1=14.1347$, while the uncorrected
$g(1-7/8)$ gives $17.85$. A better, purely asymptotic test is in the
following section.

Combining Eqs.~\eqref{eq:En} and \eqref{eq:gamman}, and expanding to first order in the
(constant) difference between the two arguments,
\begin{equation}
E_n-\gamma_n \sim g'(n)\left(\frac{\theta}{2\pi}-\frac{1}{8}\right).
\label{eq:diff1}
\end{equation}
Using $W'(x)=W(x)/\{x[1+W(x)]\}$, we get $g'(n)=2\pi/[1+W(n/e)]$, so
\begin{equation}
E_n-\gamma_n \sim \frac{\theta-\pi/4}{1+W(n/e)} .
\label{eq:diff2}
\end{equation}
For comparison, if we had (incorrectly) used $N(\gamma_n)=n$ directly, i.e.\
$\gamma_n\sim g(n-7/8)$, the same steps would give instead
\begin{equation}
E_n-\gamma_n \sim \frac{\theta-5\pi/4}{1+W(n/e)} .
\label{eq:diff2naive}
\end{equation}
For large $x$, $W(x)\sim \ln x-\ln\ln x+\dfrac{\ln\ln x}{\ln x}+O\!\big((\ln\ln x/\ln
x)^2\big)$, so $W(n/e)\sim\ln n-1-\ln(\ln n-1)$, and Eq.~\eqref{eq:diff2} becomes
\begin{equation}
E_n-\gamma_n \sim \frac{\theta-\pi/4}{\ln n-\ln(\ln n -1)} .
\label{eq:diff3}
\end{equation}
So $E_n\to\gamma_n$ as $n\to\infty$, but only very slowly, like $1/\ln n$. For $\theta=1.417\pi$ the coefficient is $\theta-\pi/4=1.167\pi=3.666$, while the naive one is $\theta-5\pi/4=0.524$.

\section{Numerical check up to $n=10^5$}
\label{sec:num}

\begin{figure}[t]
\centering
\includegraphics[width=0.9\linewidth]{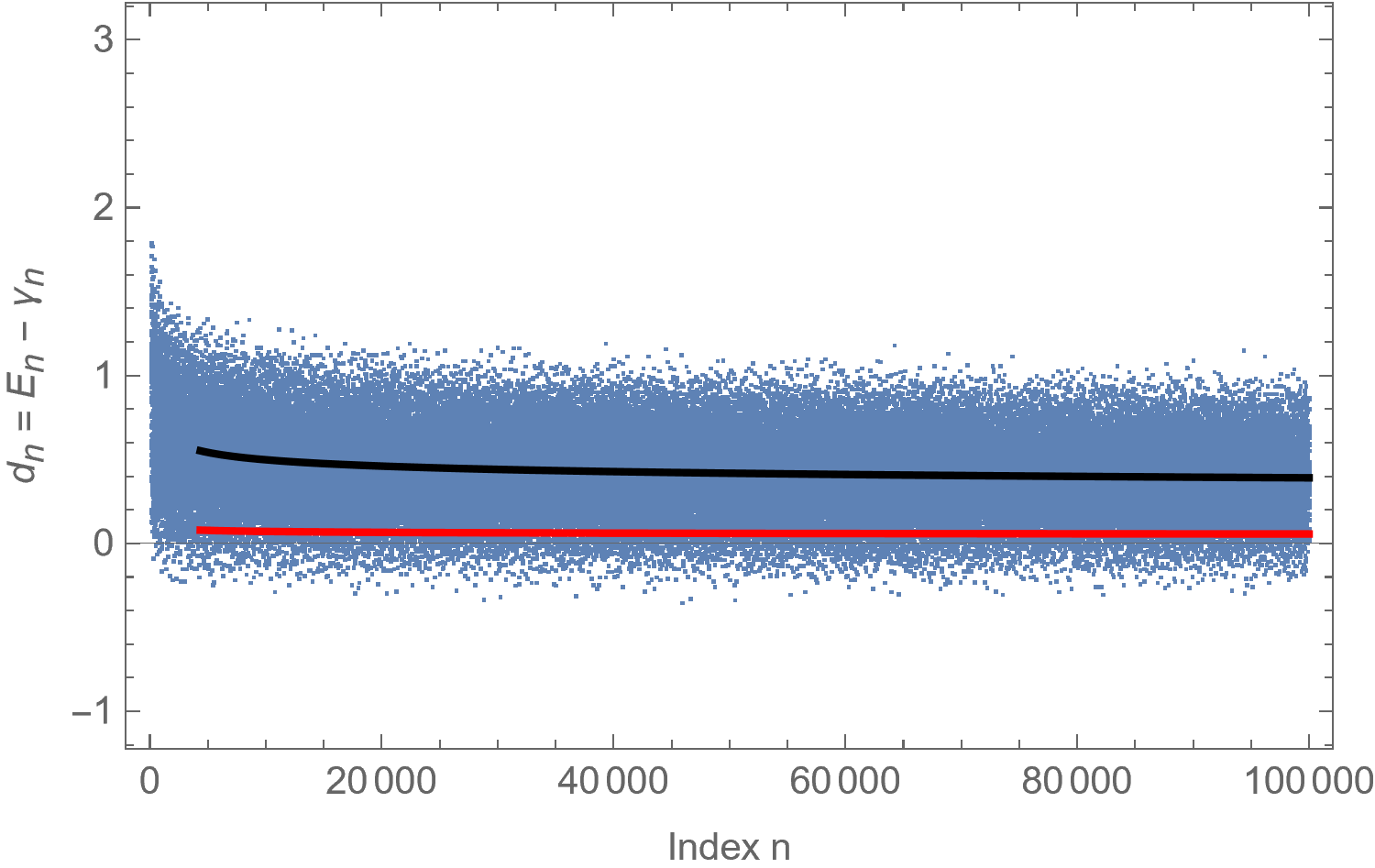}
\caption{(Color online) $d_n=E_n-\gamma_n$, with $E_n=g\left(n-\frac32+\frac{\theta}{2\pi}\right)$,
for the first $10^5$ zeta zeros (blue dots, $\theta=1.417\pi$). Black curve: Eq.~\eqref{eq:diff2},
coefficient $\theta-\pi/4$. Red curve: naive prediction, Eq.~\eqref{eq:diff2naive}, coefficient
$\theta-5\pi/4$. For $n\ge10^3$ the mean and median of $d_n[1+W(n/e)]$ are $3.666$ and
$3.667$. The scatter around the black curve comes from $S(\gamma_n)$ and is consistent with
Selberg's central limit theorem, whose variance grows like $\ln\ln T$~\cite{Selberg}.}
\label{fig:fig2}
\end{figure}

We tested Eqs.~\eqref{eq:diff2} and \eqref{eq:diff2naive} against the first $10^5$
zeta zeros tabulated by Odlyzko~\cite{Odlyzko}. Figure~\ref{fig:fig2} shows
$d_n\equiv E_n-\gamma_n$ (blue dots), with $E_n$ from
Eq.~\eqref{eq:En}, along with two curves: the black curve is the prediction of
Eq.~\eqref{eq:diff2} (coefficient $\theta-\pi/4=3.666$), and the red curve is the naive
prediction of Eq.~\eqref{eq:diff2naive} (coefficient $\theta-5\pi/4=0.524$), both with
$\theta=1.417\pi$. The red curve sits far below the point cloud, while the black curve
runs through its center. If we fit the coefficient directly to the data --- computing
$d_n\,[1+W(n/e)]$ at each point with $n\ge10^3$ and taking the mean and median --- we get
\begin{equation}
\overline{d_n[1+W(n/e)]} = 3.666\ (\text{mean}),\quad 3.667\ (\text{median}),
\end{equation}
which agrees with the black curve's coefficient $\theta-\pi/4=3.666$ to better than
$0.1\%$, and clearly disagrees with the red curve's $0.524$. This agreement has no free
parameter. The only inputs are the saddle-point phase of Sec.~\ref{sec:saddle}, the root
labeling $m=n-2$ checked against exact roots, and the midpoint treatment of $N(T)$ in
Sec.~\ref{sec:RvM}. We stress that the scatter of the blue dots is not a property of
$E_n$, which is a smooth function of $n$. It comes from $\gamma_n$: to first order,
$d_n=\text{smooth part}-S(\gamma_n)\,g'(n)$.

One spurious root near $E\approx947.5$ (originally called $E_{607}$) was found and
removed this way: we compared each level spacing $\Delta E_n$ with the leading-order
prediction $2\pi/\ln(E_n/2\pi)$. Their ratio stays close to $1$ across the whole
spectrum, except at one point near $n=607$, where it drops to $0.36$. This tells us
that the root-finder likely missed or merged two nearby roots there, rather than
finding a real level.

\section{Discussion}

We have shown that the eigenvalues of the SR self-adjoint extension problem track the
Riemann zeta zeros, with $E_n-\gamma_n\simeq(\theta-\pi/4)/[1+W(n/e)]$, which decays like
$1/\ln n$. This is confirmed numerically over five orders of magnitude in $n$. The shape
comes from the Lambert-$W$ inversion of the leading-order saddle-point phase, which
matches the rigorous asymptotic formula for $K_\nu(x)$ at large imaginary
order~\cite{Temme94,Balogh67,BST13,Tseng23}. The size comes from two integer shifts: the
$n$-th numerical root is the $(n-2)$-th root of the asymptotic condition, and the midpoint
treatment of $N(T)$ at $T=\gamma_n$ gives $\gamma_n\sim g(n-11/8)$. An independent check
of the $S(T)$-at-a-jump convention in the primary literature (e.g.\ \cite{Titchmarsh}) would
still be welcome.

Our result also shows the limit of this kind of model. The spectrum $E_n$ is a smooth
function of $n$, like the smooth part $\theta(T)/\pi+1$ of $N(T)$. It has no counterpart of
the fluctuating term $S(T)$, which carries the information on the prime numbers. Hence
$E_n$ can follow $\gamma_n$ only up to the scatter $S(\gamma_n)g'(n)$. This is consistent
with the remark of Berry and Keating~\cite{BK2011} that the absence of a connection with
the primes is shared by all variants of $xp$ (see also Ref.~\cite{Sierra12}). Note that
$\theta=\pi/4$ would remove the leading $1/\ln n$ term. We did not use this as a fit,
because $\theta$ was fixed by $\gamma_1$ and not tuned to the large-$n$ data.

\section*{Acknowledgments}
D.P. used Claude (Anthropic) as an aid in this work, including exploring the
steepest-descent calculation, searching the literature for the exact asymptotics
of $K_\nu(x)$ at large imaginary order, and developing the argument in
Sec.~\ref{sec:RvM} concerning the treatment of $N(T)$ at $T=\gamma_n$. All
derivations, numerical results, and conclusions were independently verified by
the authors, who take full responsibility for the content of this paper.


\begin{thebibliography}{99}
\bibitem{BK} M. V. Berry and J. P. Keating, \emph{$H=xp$ and the Riemann Zeros}, in
\emph{Supersymmetry and Trace Formulae: Chaos and Disorder}, edited by I. V. Lerner,
J. P. Keating, and D. E. Khmelnitskii (Plenum, New York, 1999).
\bibitem{BK2011} M. V. Berry and J. P. Keating, \emph{A compact hamiltonian with the same asymptotic mean spectral density as the Riemann zeros}, J. Phys. A: Math. Theor.  \textbf{44}, 285203 (2011).
\bibitem{Sierra12} G. Sierra, \emph{General covariant $xp$ models and the Riemann zeros},  J. Phys. A: Math. Theor.  \textbf{45}, 055209 (2012); arXiv:1110.3203 (math-ph).
\bibitem{SR} G. Sierra and J. Rodr\'iguez-Laguna, \emph{$H=xp$ Model Revisited and the
Riemann Zeros}, Phys. Rev. Lett. \textbf{106}, 200201 (2011); arXiv:1102.5356 (math-ph).
\bibitem{Temme94} N. M. Temme, \emph{Steepest descent paths for integrals defining the
modified Bessel functions of imaginary order}, Methods Appl. Anal. \textbf{1}, 14
(1994).
\bibitem{Balogh67} C. B. Balogh, \emph{Asymptotic expansions of the modified Bessel
function of the third kind of imaginary order}, SIAM J. Appl. Math. \textbf{15}, 1315
(1967).
\bibitem{BST13} A. R. Booker, A. Str\"ombergsson, and H. Then, \emph{Bounds and
algorithms for the K-Bessel function of imaginary order}, LMS J. Comput. Math.
\textbf{16}, 78 (2013).
\bibitem{Tseng23} J. Tseng, \emph{An asymptotic for the K-Bessel function using the
saddle-point method}, arXiv:2302.09962 (math.CA).
\bibitem{Edwards} H. M. Edwards, \emph{Riemann's Zeta Function} (Academic Press, New
York, 1974).
\bibitem{Titchmarsh} E. C. Titchmarsh, \emph{The Theory of the Riemann Zeta-Function},
2nd ed., revised by D. R. Heath-Brown (Oxford University Press, 1987).
\bibitem{Littlewood} J. E. Littlewood, \emph{On the zeros of the Riemann zeta-function},
Proc. Cambridge Philos. Soc. \textbf{22}, 295 (1924).
\bibitem{Odlyzko} A. M. Odlyzko, \emph{Tables of zeros of the Riemann zeta function},
\url{https://www-users.cse.umn.edu/~odlyzko/zeta_tables/index.html}.
\bibitem{Selberg} A. Selberg, \emph{On the remainder in the formula for $N(T)$, the
number of zeros of $\zeta(s)$ in the strip $0<t<T$}, Avh. Norske Vid. Akad. Oslo
\textbf{1}, 1 (1944).
\end{thebibliography}
\end{document}